# Optimized Strategies for Peak Shaving and BESS Efficiency Enhancement through Cycle-Based Control and Cluster-Level Power Allocation

***Guo Gan*, Li Junhui, Mu Gang, Yan Gangui***

*Key Laboratory of Modern Power System Simulation and Control & Renewable Energy Technology, Ministry of Education (Northeast Electric Power University), Jilin 132012, China.*

**guogan@outlook.com*

**Abstract**

Battery Energy Storage Systems (BESS) are essential for peak shaving, balancing power supply and demand while enhancing grid efficiency. This study proposes a cycle-based control strategy for charging and discharging, which optimizes capture rate (CR), release rate (RR), and capacity utilization rate (CUR), improving BESS performance. Compared to traditional day-ahead methods, the cycle-based approach enhances operational accuracy and reduces capacity waste, achieving a CUR increase from 75.1% to 79.9%. An innovative cluster-level power allocation method, leveraging an improved Particle Swarm Optimization (PSO) algorithm, is introduced. This strategy reduces daily energy loss by 174.21 kWh (3.7%) and increases BESS efficiency by 0.4%. Transient and steady-state energy loss components are analyzed, revealing that transient loss proportion decreases significantly as power depth increases, from 27.2% at 1 MW to 1.3% at 10 MW. Simulations based on a detailed Simulink/Simscape model validate these methods, demonstrating enhanced peak shaving effectiveness and prolonged BESS lifespan by reducing equivalent cycles. The study provides a robust framework for optimizing BESS performance and efficiency in real-world applications.

# 0 Introduction

Energy storage systems (ESS) play a crucial role in managing the balance between power supply and demand, particularly in the context of peak shaving. Peak shaving involves reducing the peak demand on power systems, which can alleviate stress on the grid and improve overall efficiency. The integration of renewable energy sources, such as wind and solar, has increased the need for effective peak shaving strategies due to their intermittent nature[1][2][9].

The efficiency of energy storage systems is a critical factor in their economic feasibility and operational effectiveness. Battery energy storage systems (BESS) are commonly used for peak shaving, storing energy during off-peak hours and discharging during peak demand periods. Recent advancements have focused on improving the service life and efficiency of BESS by using power electronics to manage power flow and accommodate batteries of varying ages and technologies[3]. Additionally, flywheel energy storage systems have been explored for their ability to provide peak shaving services with minimal energy losses[6].

Various strategies have been developed to optimize the participation of energy storage in peak shaving. These include prioritizing energy storage over traditional thermal generators for peak shaving, which has been shown to improve the utilization of energy storage systems[1]. Moreover, the implementation of optimal capacity and power configurations for grid-side energy storage can enhance power stability and reduce peak regulation pressure[2]. Advanced algorithms and control strategies have been proposed to determine the optimal size and location of BESS, as well as to schedule their operation for maximum efficiency[4-5].

The economic viability of energy storage systems in peak shaving applications is influenced by factors such as system efficiency, cost savings, and market trading mechanisms.[7] Integrated systems that combine energy storage with other technologies, such as liquid air energy storage and combined cycle power plants, have been proposed to maximize efficiency and economic performance[8]. Additionally, market trading mechanisms that involve energy storage participation in multi-source peak shaving can facilitate the integration of renewable energy sources and improve grid stability[10].

Energy storage systems are essential for effective peak shaving in power systems, particularly with the increasing penetration of renewable energy sources. Advances in technology and strategic operation models have improved the efficiency and economic feasibility of these systems. By optimizing the size, location, and operation of BESS, and integrating them with other technologies, the power grid can achieve more stable and efficient peak shaving. These developments highlight the importance of continued research and innovation in energy storage technologies and their applications in power systems. Fig. 1 illustrates the research structure of this study.

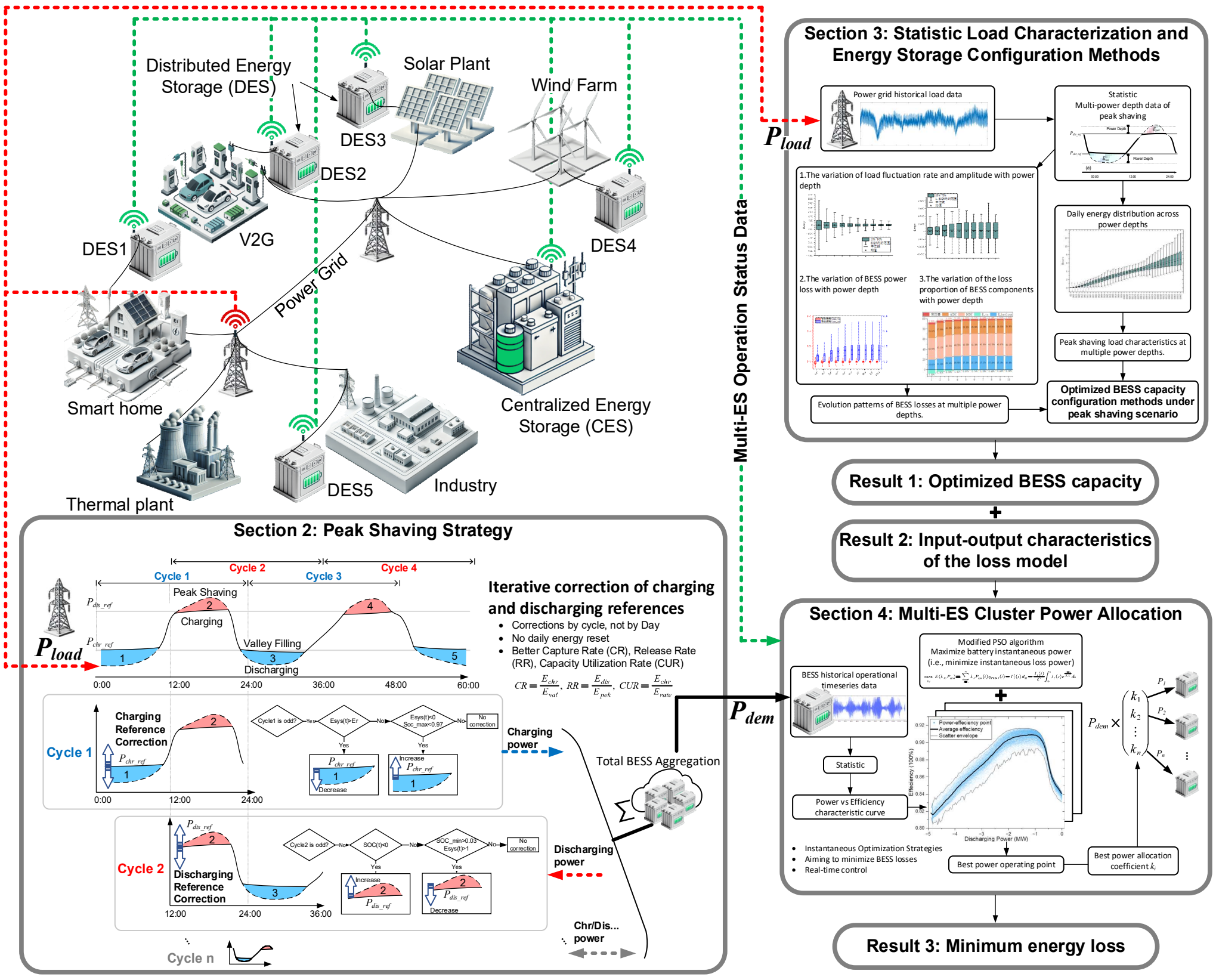

Fig. 1 Research structure overview

# 1 BESS energy loss model

## 1.1 BESS simulation

A 5MW/20MWh BESS model was developed base on the Simulink/Simscape platform, comprising 100 battery clusters (50 kW/200 kWh each) connected in parallel via PCS on the AC side, as shown in Fig. 2. A 5 MW/20 MWh BESS model was developed on the Simulink/Simscape platform, comprising 100 battery clusters (50 kW/200 kWh each) connected in parallel via PCS on the AC side, as shown in Fig. 2。

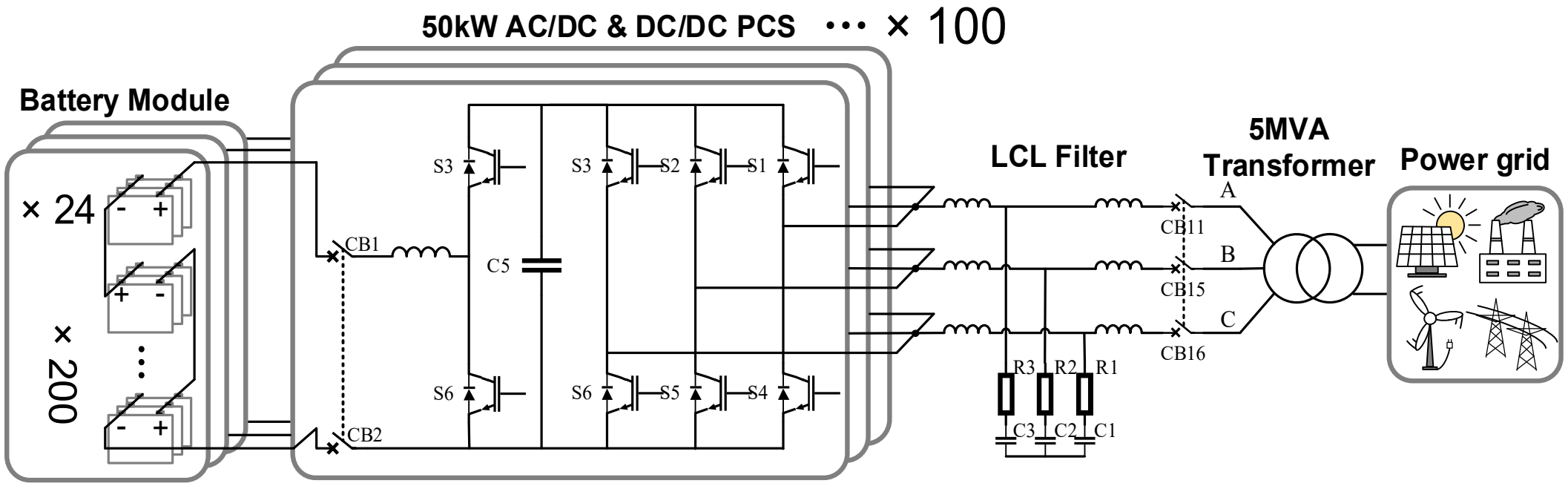

Fig. 2 BESS circuit structure

As shown in Fig. 2, the BESS comprises 100 battery clusters, each with a rated power of 50 kW. The low-voltage DC side of the 50 kW DC/DC PCS in each cluster connects to the battery module, while the high-voltage DC side interfaces with the DC side of a three-phase bridge AC/DC converter at a constant voltage of 700 V, forming a single battery cluster. As shown in Fig. 2, the BESS comprises 100 battery clusters, each with a rated power of 50 kW. The low-voltage DC side of the 50 kW DC/DC PCS in each cluster connects to the battery module, while the high-voltage DC side interfaces with the DC side of a three-phase bridge AC/DC converter at a constant voltage of 700 V, forming a single battery cluster.

In the BESS model, each lithium battery cell has a capacity of 12.5 Ah, with its equivalent circuit model consistent with the description in Section 1.4. The parameters are $R_{\Omega}$=0.0232Ω，$R_{\mathrm{p}}$=0.0185Ω，$C_{\mathrm{p}}$=12091F。Each battery module comprises 24 cells in parallel and 200 groups in series.

## 1.2 Transformer loss model

Based on the T-type equivalent circuit model of the transformer, its losses primarily originate from no-load losses (core losses) and load losses (copper losses). The relationship between transformer loss power and load factor $\lambda$ is expressed as:

$$\begin{aligned} P_{TF_loss}(\lambda) &= P_{Fe} + P_{Cu} \\ &= P_{Fe} + \lambda^2 P_{kN} \end{aligned} \tag{1}$$

## 1.3 Converter loss model

Based on empirical data from the Hongsheng PWS1 50k PCS, this study derives a fitted relationship between efficiency of the PCS and load factor $\lambda$(2)：

$$\eta_{PCS}(\lambda) = \sum_{i=0}^{4} \alpha_i \lambda^i \tag{2}$$

In equation: $\alpha_0$=0.7868，$\alpha_1$=0.7955，$\alpha_2$=-2.073，$\alpha_3$=2.137，$\alpha_4$=-0.8137

## 1.4 Battery loss model

### 1.4.1 Battery circuit model

Currently, the nonlinear battery model with an RC circuit is widely used in traction batteries and large-scale renewable energy storage systems. This model captures the fundamental dynamic and time-varying electrical characteristics of batteries while balancing computational complexity and accuracy. Therefore, this study adopts a first-order RC Thevenin equivalent model as the circuit model for individual battery cells:

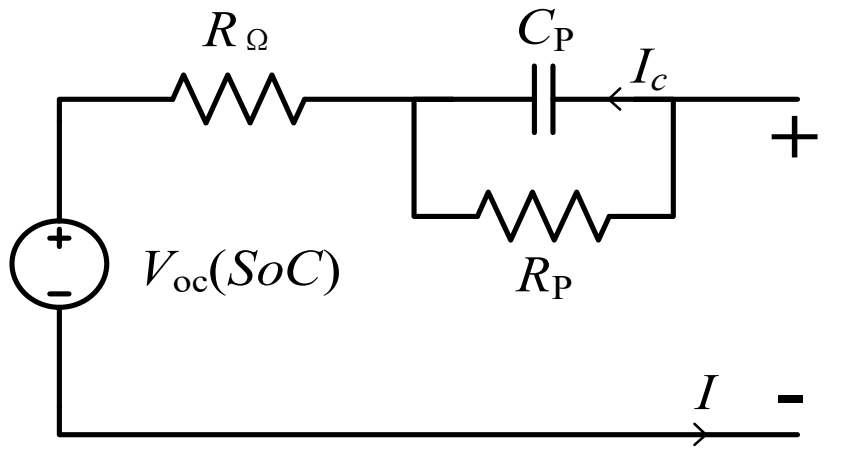

Fig. 3 Battery Equivalent Circuit Model

The $V_{oc}$ of the battery exhibits a nonlinear functional relationship with the SoC, which can be approximated using polynomial fitting:

$$V_{oc}(SoC) = \sum_{i=0}^{3} \beta_i SoC^i \tag{3}$$

Pulse discharge experiments were conducted on multiple Shenhang LR12500SA lithium batteries, and the averaged results determined the following coefficients: $\beta_0$=2.484，$\beta_1$=2.608，$\beta_2$=-5.252，$\beta_3$=3.603 (units: V).

### 1.4.2 Steady-state loss power component $P_{bat_ss}$

Battery energy loss primarily originates from ohmic resistance and polarization resistance. Ohmic resistance characterizes the internal material resistance and interfacial contact resistance within the battery. The polarization resistance, together with the polarization capacitance, forms an RC circuit that models the electrical behavior of polarization effects, where the polarization resistance represents the impedance caused by polarization to the movement of internal charges. The total loss power of the battery $P_{bat_loss}$, is the sum of ohmic resistance loss $P_{bat_rint}$ and polarization resistance loss $P_{bat_rp}$.

$$\begin{aligned} P_{bat_loss} &= P_{bat_rint} + P_{bat_rp} \\ &= RI^2 + R_p I_P^2 \end{aligned} \tag{4}$$

The relationship between the polarization current $I_p$ and the terminal current $I$ is expressed as:

$$I_p = \frac{1}{R_P C} e^{\frac{-t}{R_P C}} \int_0^t I(s) e^{\frac{s}{R_P C}} ds = I - \int_0^t I'(s) e^{\frac{s-t}{R_P C}} ds \tag{5}$$

Substituting $I_p$ into Equation (4) yields:

$$P_{bat_loss} = RI^2 + R_p I^2 + R_p \left( \int_0^t I'(s) e^{\frac{s-t}{R_P C}} ds \right)^2 - 2R_p I \int_0^t I'(s) e^{\frac{s-t}{R_P C}} ds \tag{6}$$

From the above expression, it can be concluded that, as the battery model includes an RC transient circuit, the battery loss power can be divided into steady-state and transient components. The steady-state loss power component, $P_{bat_ss}$, is extracted and expressed as:

$$\begin{aligned} P_{bat_ss}(I) &= P_{bat_rint}(I) + I^2 R_P \\ &= I^2 (R + R_P) \end{aligned} \tag{7}$$

From Equation (7), it is evident that $P_{bat_ss}$ is always positive. The steady-state loss power component represents the power loss caused by the load current passing through the internal resistance of the battery, and its magnitude is solely determined by the terminal current $I$.

### 1.4.3 Transient-state loss power component $P_{bat_ts}$

The transient loss power component of the battery, denoted as $P_{bat_ts}$, is extracted as follows:

$$\begin{aligned} P_{bat_ts}(t,I,I') &= P_{bat_rp}(t,I,I') - I^2 R_P \\ &= R_p \left( \int_0^t I'(s) e^{\frac{s-t}{R_P C}} ds \right)^2 - 2R_p I \int_0^t I'(s) e^{\frac{s-t}{R_P C}} ds \end{aligned} \tag{8}$$

In the equation, the decay speed of the integral part is determined by the time constant $R_p C$.。

# 2 BESS peak shaiving strategy

## 2.1 Charging/discharging power

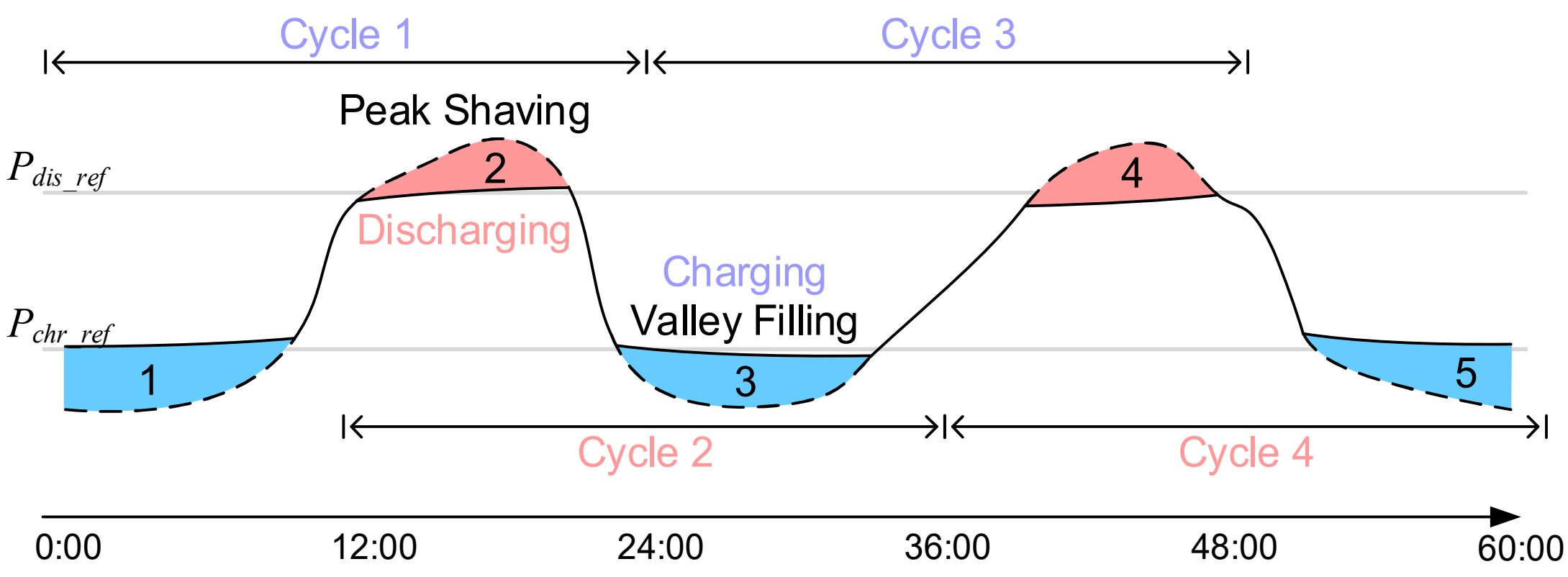

Fig. 4 BESS peak load regulation procedure

The peak shaving power depth $P_d$ is defined as the difference between the maximum (or minimum) load and the peak shaving target $P_{dis_ref}$ (or $P_{chr_ref}$), reflecting the peak shaving intensity of BESS. Generally, the power depth $P_d$ is equal to the rated power $P_r$ of BESS to maximize the power regulation ability, thereby improving the peak shaving effect.

During the operation of the BESS, its peak shaving ability is constrained both by rated power and rated capacity. Therefore, the formulation of the peak shaving plan should first forecast the day-ahead load curve, reasonably segment the peak shaving cycles, and determine the control targets $P_{chr_ref}$ and $P_{dis_ref}$ for the peak and valley loads within each cycle. On this basis, the charging and discharging power of BESS in the $i$-th peak shaving cycle can be defined as:

$$P_{chr}(cyc\ i)=\begin{cases} P_r & P_{load}(t)\in(0,P_{ref_chr}-P_r] \quad \& \ SoC\in[0,1) \\ P_{ref_chr}-P_{load}(t) & P_{load}(t)\in(P_{ref_chr}-P_r,P_{ref_chr}) \ \& \ SoC\in[0,1) \\ 0 & other \end{cases} \tag{9}$$

$$P_{dis}(cyc\ i)=\begin{cases} P_{ref_dis}-P_{load}(t) & P_{load}(t)\in(P_{ref_dis},P_{ref_dis}+P_r) \ \& \ SoC\in(0,1] \\ -P_r & P_{load}(t)\in[P_{ref_dis}+P_r,+\infty) \ \& \ SoC\in(0,1] \\ 0 & other \end{cases} \tag{10}$$

$P_{chr}(cyc\ i)$ and $P_{dis}(cyc\ i)$ are the energy storage charging and discharging powers at time $t$.

As mention above, the setting of the charging and discharging power of the energy storage system is jointly determined by the load $P_{load}$ and the SoC level. The entire process is illustrated by Fig. 5.

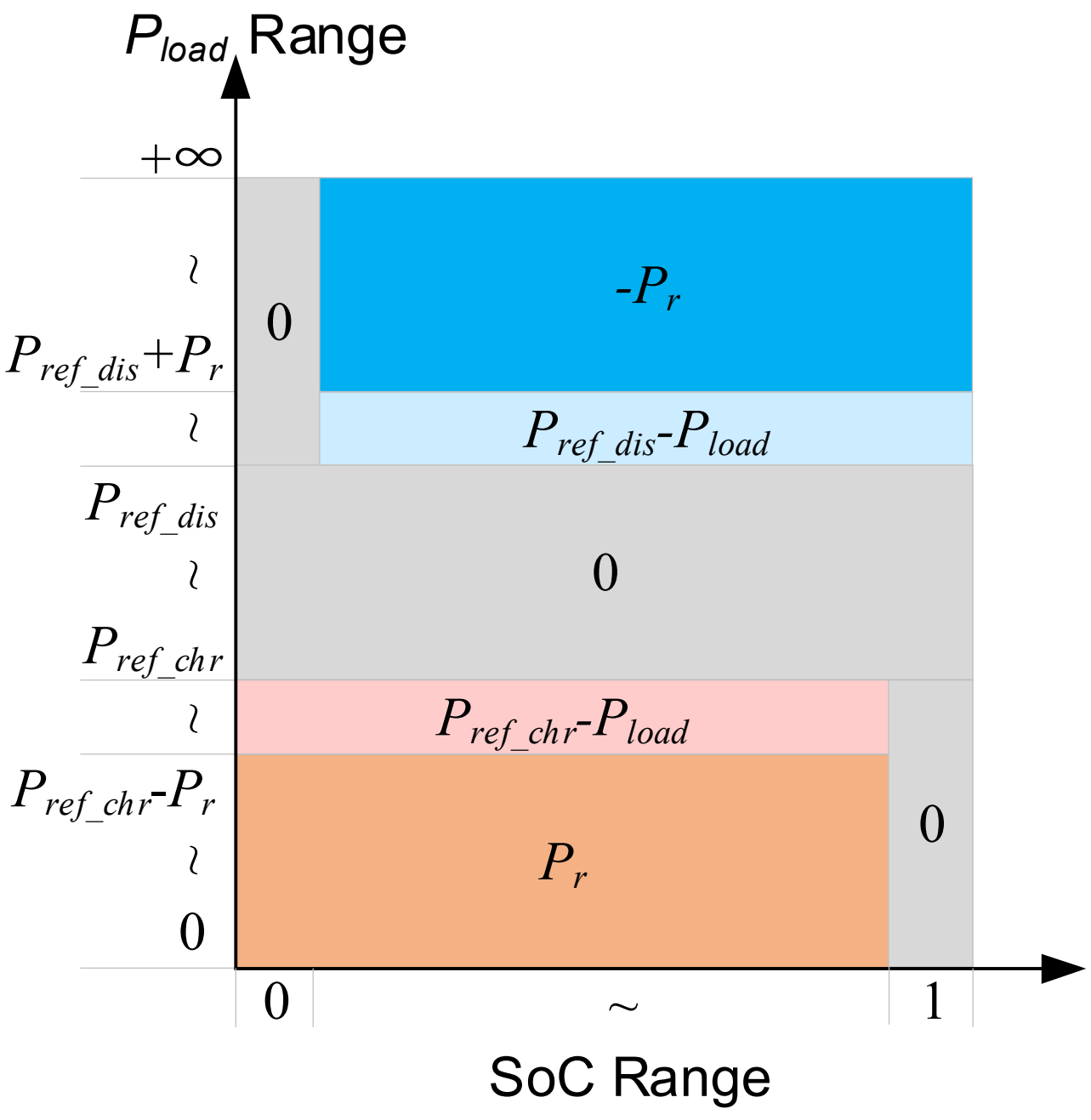

Fig. 5 BESS charging and discharging power setting rules.

By summing all $n$ peak shaving cycles over the entire time domain, the total charging and discharging power demand of the ESS over the full time domain, $P_{sys}(t)$, can be obtained:

$$\begin{aligned} P_{sys}(t) &= \sum_{i=1}^{n} P_{chr}(cyc\ i)+P_{dis}(cyc\ i) \\ &= \sum_{i=1}^{n} P_{dem}(cyc\ i) \end{aligned} \tag{11}$$

$P_{sys}(t)$ represents the ESS power demand, with positive values for charging and negative for discharging.

## 2.2 Charging/discharging reference correction

If the constraint is not met, it is necessary to adjust the peak-valley control target reference $P_{chr_ref}$ and $P_{dis_ref}$. The adjustment process is shown in Fig. 6:

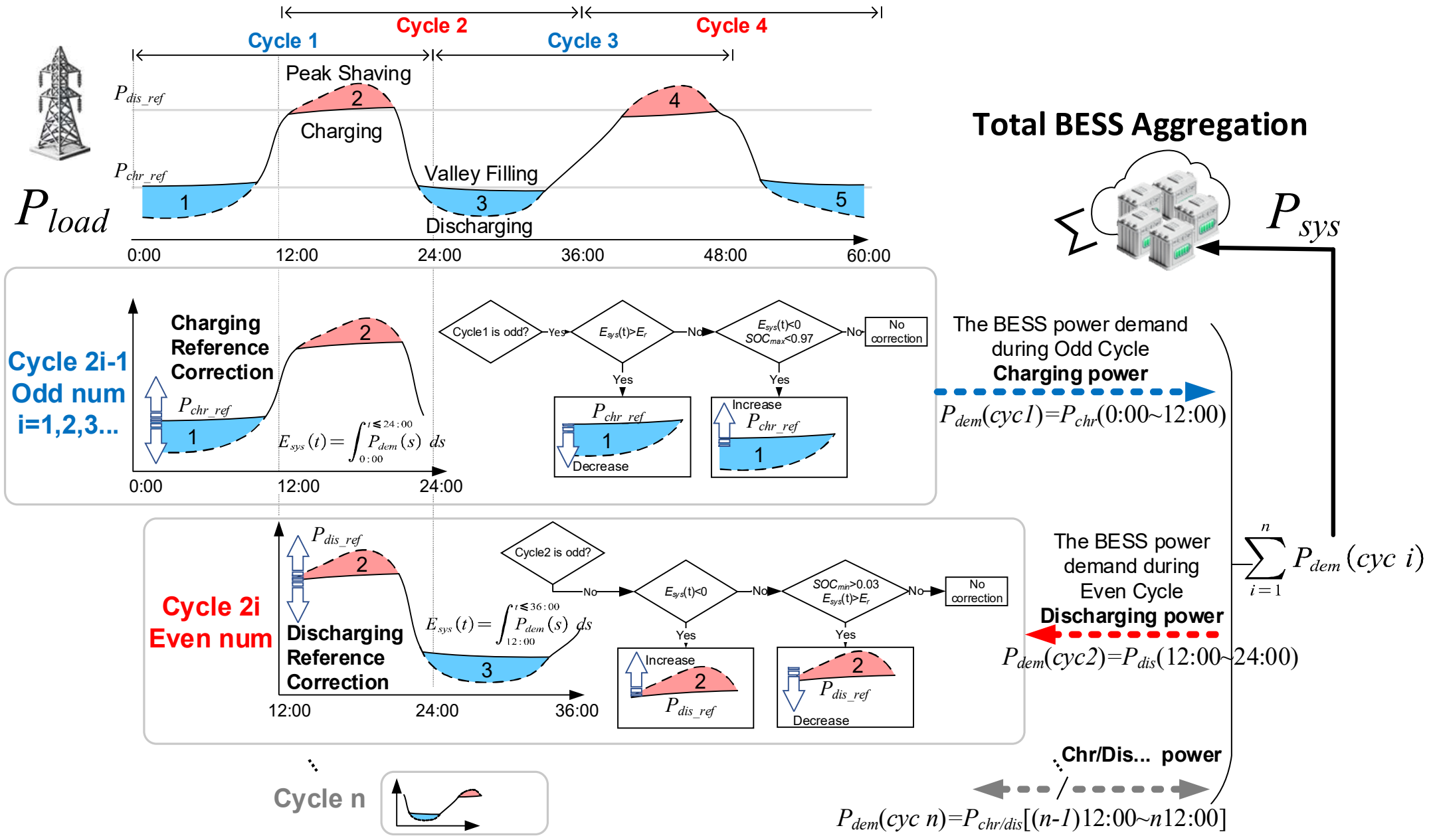

Fig. 6 Correction principle of charging and discharging reference limits.

As shown in Fig. 6, the peak shaving process generally consists of alternating charging and discharging intervals. The charging/discharging intervals are determined based on the forecasted load, and adjacent charging/discharging intervals form a charging/discharging cycle. Each cycle is indexed accordingly, i.e., Charging Interval 1 and Discharging Interval 2 form Cycle 1, Discharging Interval 2 and Charging Interval 3 form Cycle 2, and so on. The reference limits of the previous charging or discharging interval are adjusted based on the next interval within the same cycle.

The accumulated energy of the BESS at time $t$ in each peak shaving cycle is expressed as $E_{cyc}(t)$:

$$E_{cyc}(t)=\int_{0}^{t} P_{dem}(s)\,ds \tag{12}$$

The maximum and minimum energy values of $E_{cyc}(t)$ within a cycle correspond to the SoC values defined as $SOC_{max}$ and $SOC_{min}$ respectively. Based on this, the correction rules for the charging and discharging reference within the cycle are presented in Table 1.

Table 1 $P_{chr_ref}$ and $P_{dis_ref}$ correction method conditon and rules

| Condition | | | Action |
|---|---|---|---|
| Cycle number is odd | $E_{cyc}(t)> E_r$ | $P_{chr_ref}<P_{dis_ref}$ and n_chr1<100 | Over charge, reduce $P_{chr_ref}$ |
| | $E_{cyc}(t)<0$, $t\in[0,\infty)$ and $SOC_{max}<0.97$ | | Under charge, increase $P_{chr_ref}$ |
| Cycle number is even | $E_{cyc}(t)>E_r$, $t\in[0,\infty)$ and $SOC_{min}>0.03$ | $P_{chr_ref}<P_{dis_ref}$ and n_dis4<100 | Under discharge, reduce $P_{dis_ref}$ |
| | $E_{cyc}(t)<0$ | | Over discharge, increase $P_{dis_ref}$ |

In the table, n_chr1 and n_dis4 represent the maximum consecutive correction iterations in the charging and discharging adjustment procedures, respectively, which are temporarily set to 100.

Compared with traditional peak shaving methods, the process can be represented by the flowchart shown in Fig. 7.

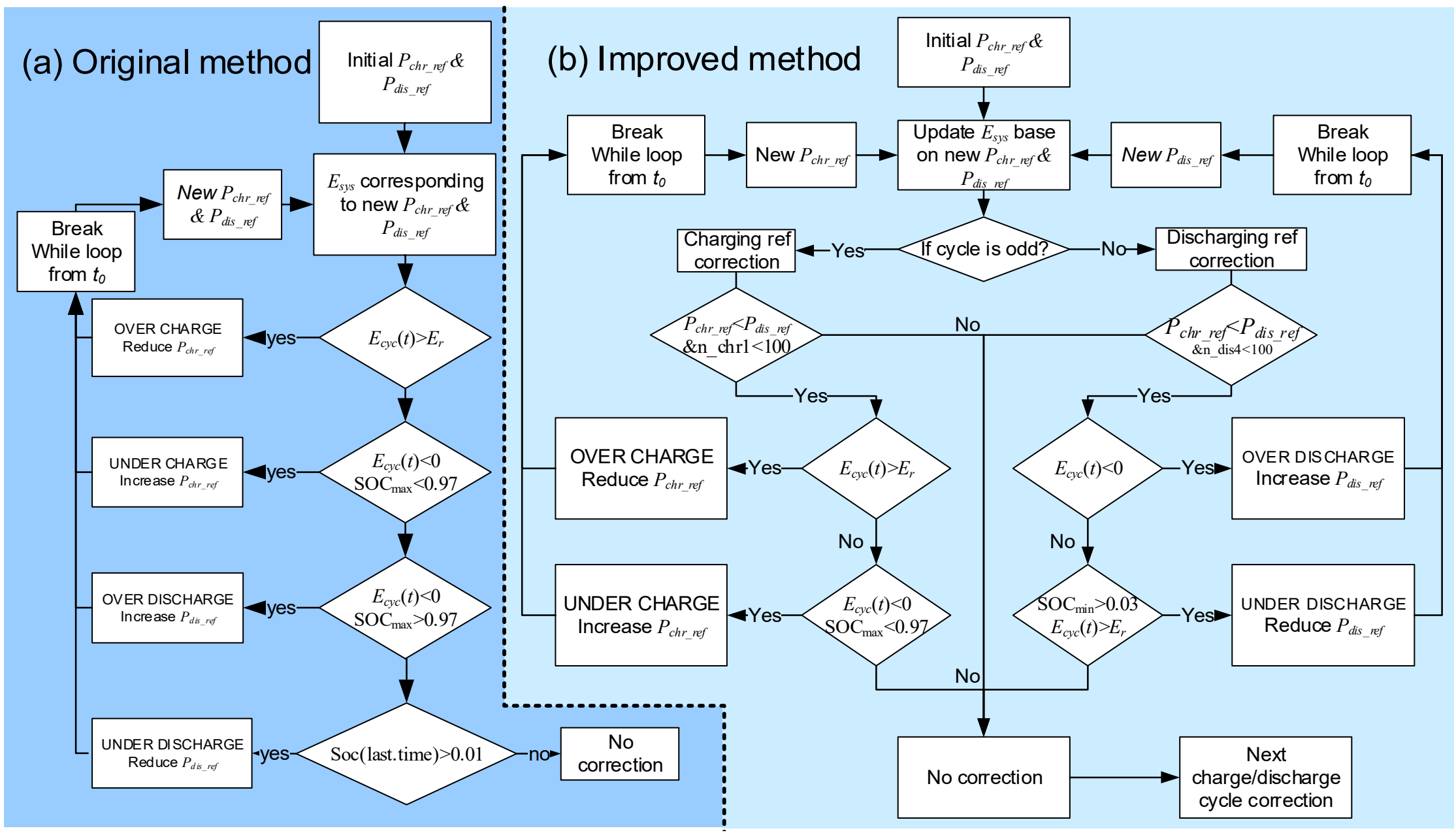

Fig. 7 Comparison of the original and improved BESS peak shaving charge/discharge reference correction processes.

After comparison, it is found that the improved method provides a more refined distinction of operating conditions. The correction of charging and discharging references is performed by cycle, not by day. Although the process is more complex, each iteration involves fewer steps, so the computational burden does not increase. Moreover, the improved method enhances the peak shaving performance.

## 2.3 Performance metrics of BESS peak shaving

The improved peak shaving strategy can improve some key BESS operating metrics and optimize peak shaving performance metrics. The following section defines these indicators and analyzes the optimization effects.

As shown in Fig. 8, $E_{val}$ and $E_{pek}$ represent the valley and peak load energy at the corresponding power depth, while $E_{chr}$ and $E_{dis}$ denote the actual charging and discharging energy during the ESS peak shaving process. Their significance in the load profile can be illustrated by the shaded areas corresponding to load energy.

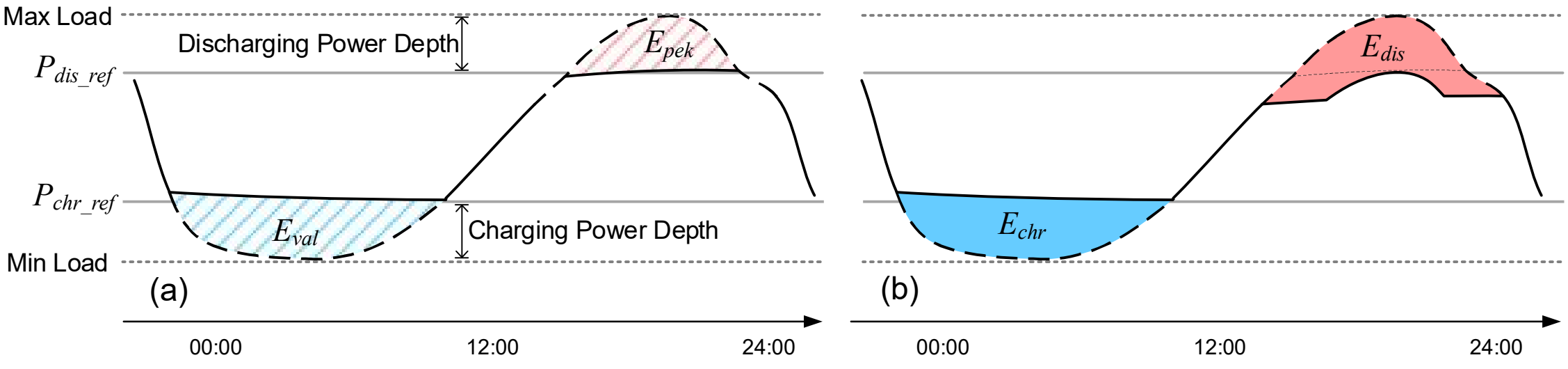

Fig. 8 Valley load energy $E_{val}$ & peak energy $E_{pek}$, and ESS charging/discharging energy $E_{chr}$&$E_{dis}$ at the corresponding power depth

To evaluate the effectiveness of the peak shaving method, this study defines the Capture Rate (CR) and Release Rate (RR) to quantify the performance of BESS in peak shaving. Additionally, the Capacity Utilization Rate (CUR) is introduced to assess the utilization level of BESS capacity in peak shaving scenarios. Their definitions are given as follows:

$$
\begin{cases}
CR = \dfrac{E_{chr}}{E_{val}} \\
RR = \dfrac{E_{dis}}{E_{pek}} \\
CUR = \dfrac{E_{chr}}{E_{rate}}
\end{cases} \tag{13}
$$

The closer these three metrics are to 1, the better.

This study analyzes the daily charging and discharging energy of a 5MW/10MWh BESS under both the original and improved peak shaving strategies over a 365-day simulation period. The results are recorded in Fig. 9, where:

- The light blue represents the improved method.
- The red represents the original method.
- The envelope lines indicate the distribution range of peak and valley energy at the corresponding power depth.

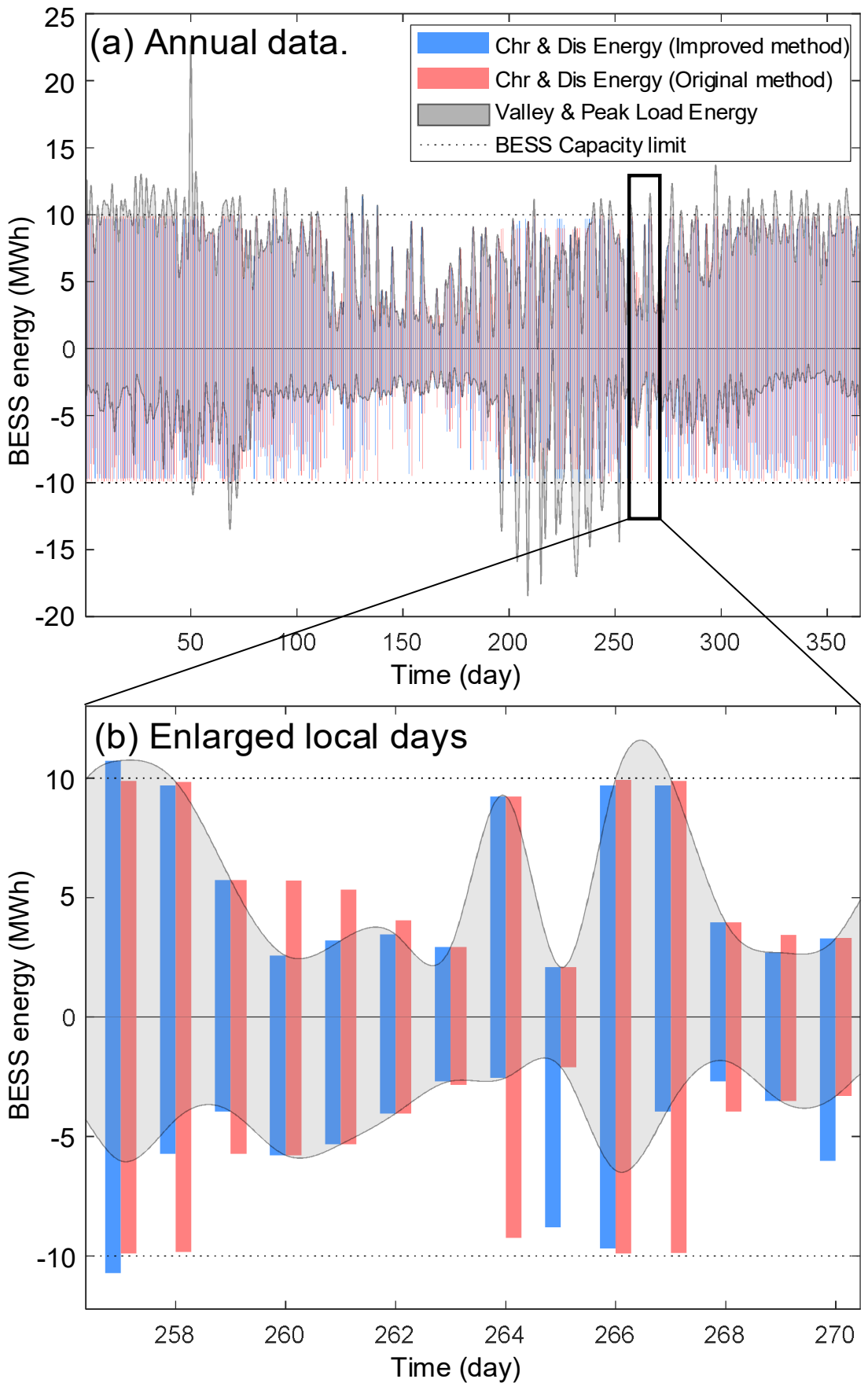

Fig. 9 Distribution of BESS charging-discharging energy and peak-valley load energy

From Fig. 9, it can be observed that the improved method does not necessarily maintain a symmetric daily charging and discharging pattern, making it more adaptive to the actual peak and valley energy distribution of the load. In contrast, the original method exhibits capacity wastage.

While time-series data provide insights into operational details, they have limitations in observing overall trends. Therefore, based on the annual simulation data in Fig. 9, the annual average values of the evaluation metrics defined in Equation (13), along with other key indicators, are statistically analyzed. The specific parameters are presented in the following table.

Table 2 Evaluation indicators for peak shaving effectiveness and BESS utilization level

| Statistical indicators | Original method | Improved method |
|---|---|---|
| Average CR | 1.0656 | 0.99778 |
| Average RR | 2.1306 | 2.0528 |
| Average CUR | 0.75104 | 0.79899 |
| Average power utilization rate | 0.9999 | 0.99712 |
| equivalent cycles | 273.9214 | 266.0955 |

Since CR and RR approaching 1 indicate an optimal peak shaving effect under the given ESS configuration, values less than 1 suggest insufficient peak shaving, while values greater than 1 indicate over-adjustment. As shown in Table 2, the improved method achieves CR and RR values closer to 1 compared to the original method.

The improved method maintains a similar average power utilization rate to the original method while achieving a higher average CUR, indicating better BESS capacity utilization while meeting peak shaving objectives.

In summary, the improved method reduces the number of equivalent full charge-discharge cycles by 7.83 over the 365-day peak shaving operation compared to the original method, effectively extending the BESS lifespan..

# 3 Correlation analysis of load and power loss data with peak shaving power depth

## 3.1 Load energy characteristics

Different peak shaving power depths represent varying peak shaving intensities, and load characteristics differ across power depths. Therefore, statistical analysis at a single depth cannot fully capture the impact of load power characteristics on BESS efficiency. It is necessary to analyze the trend of battery losses with power depth variations and establish the relationship between load characteristics and battery state to reveal both the unique and common patterns of battery losses under different operating conditions.

This section conducts a statistical analysis of battery cluster data across multiple peak shaving power depths ranging from 1 MW to 10 MW and examines the evolution of these data trends.

Before conducting the data analysis, it is important to emphasize that as the power depth increases, the number of battery clusters in the BESS also increases proportionally. For example, at a 5 MW power depth, there are 100 battery clusters (each with a rated power of 50 kW), whereas at a 10 MW power depth, there are 200 clusters. Thus, although the data correspond to different peak shaving power depths, the maximum power shared by each battery cluster remains the same at 50kW. Therefore, the differences in statistical data across power depths reflect variations in load characteristics rather than the scale of input power.

Since the load fluctuates differently each day throughout the year, the peak-valley energy distribution at each power depth forms a range. As scatter plots are difficult to interpret, key statistical characteristics such as the maximum, minimum, and median values of these distributions are summarized and presented in the box plot shown in Fig. 10.

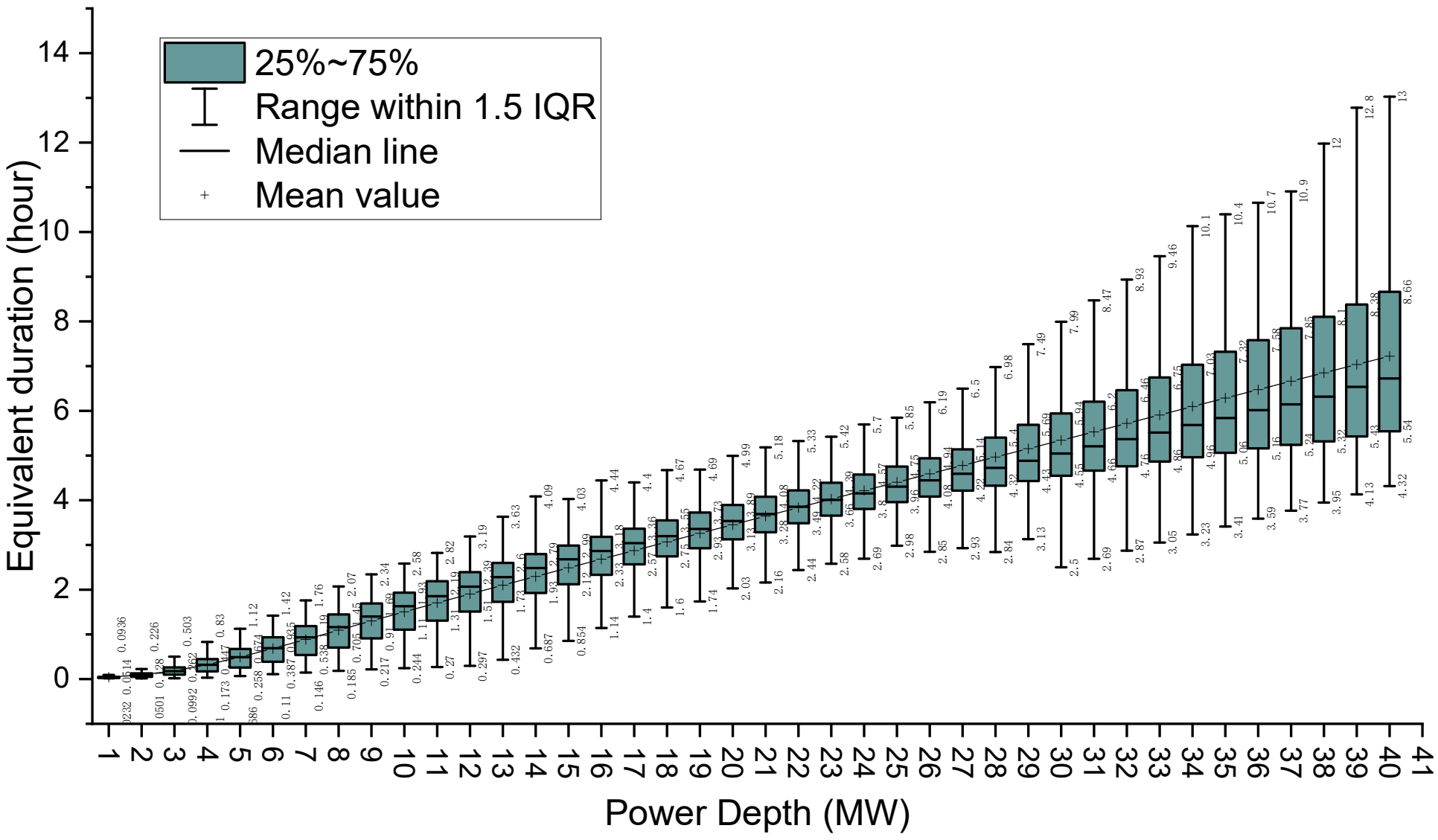

Fig. 10 Box plot of load peak-valley energy distribution at different power depths.

This figure provides insights into the energy contained in the peak and valley load values at different peak shaving power depths, which can be used as a reference for guiding the power and capacity configuration of the ESS.

## 3.2 Distribution evolution of battery port power $P_{clu}$

The distribution evolution of the battery cluster port power $P_{clu}$ and its rate of change $dP_{clu}/dt$ across different peak shaving power depths (1MW to 10MW) was analyzed. As mentioned in the previous section, the total number of battery clusters in the BESS increases proportionally with power depth. Consequently, the maximum charge and discharge power allocated to each battery cluster remains the same, effectively normalized.

During the statistical analysis, idle charging/discharging states with zero power were excluded, and only the operating battery cluster port power was considered. The horizontal axis represents different power depths, while the vertical axis corresponds to the statistical values. The results are illustrated in Fig. 11

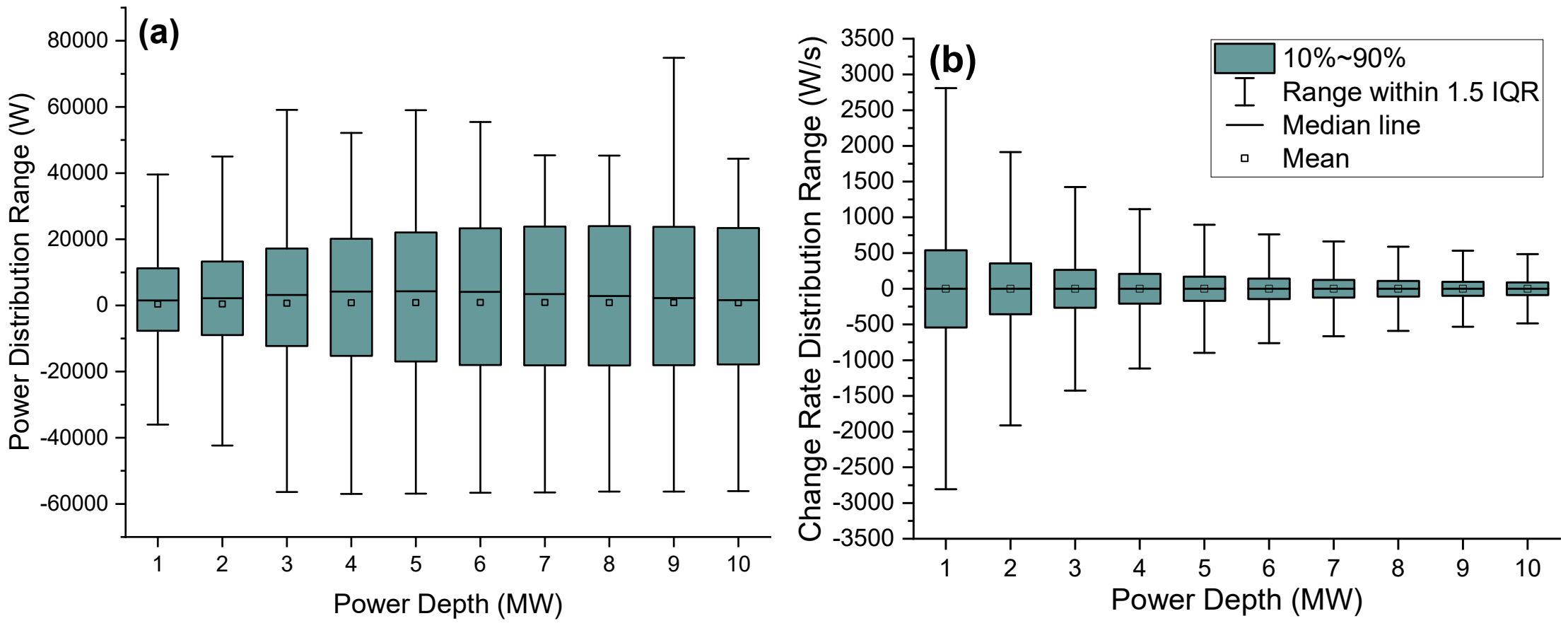

Fig. 11 battery terminal data distribution at the at different power depths (a) Power amplitude ($P_{clu}$); (b) Short - term fluctuations ($dP_{clu}/dt$)

As shown in Fig. 11(a), the magnitude of the battery cluster port power $P_{clu}$ initially increases with power depth and then stabilizes. This is because, at low depths, the load power has a lower "fullness," while at higher power depths, the base portion of the load is relatively higher in comparison to the peak, and it tends to stabilize as the depth increases.

As shown in Fig. 11(b), the distribution range of $dP_{clu}/dt$ is approximately symmetrically centered around 0, indicating that the load's increase and decrease trends are fairly balanced, and the rates of change in load are generally uniformly distributed. As the power depth increases, the distribution range gradually converges toward 0.

Since $dP_{clu}/dt$ measures short-term volatility, this suggests that with the power depth increases, the short-term volatility of the battery port power $P_{clu}$ decreases relatively. This could be due to the fact that as the power depth becomes greater, the fluctuating portion of the load decreases, while the proportion of the more stable base load increases.

## 3.3 Distribution evolution of the battery power loss $P_{loss}$

The battery loss data under different peak shaving depths is analyzed. The distribution of total battery power loss $P_{loss}$ is as follows::

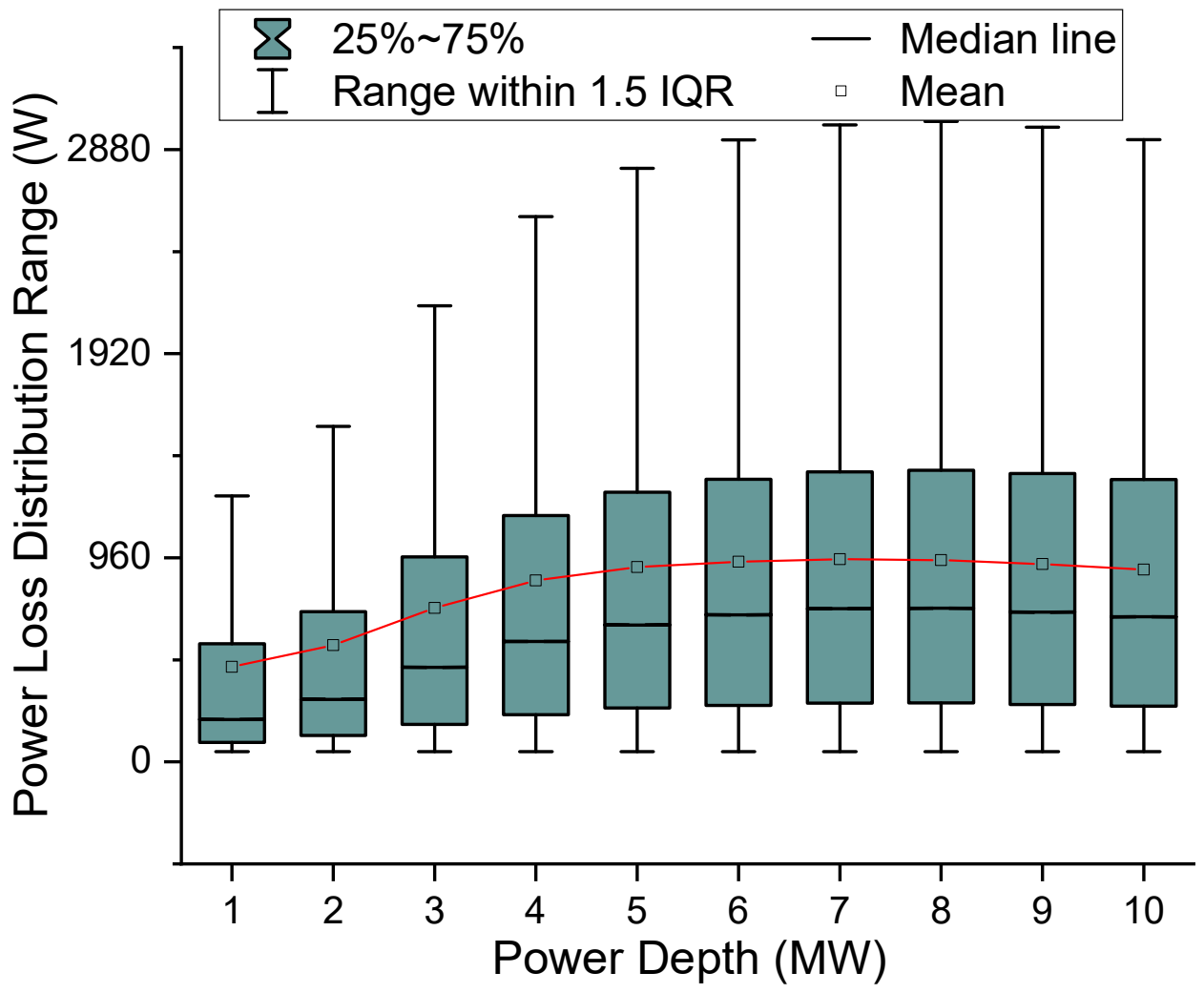

Fig. 12  Boxplot of battery power loss distribution.

The total power loss $P_{loss}$ of a single battery cluster increases gradually with the power depth and stabilizes after reaching a depth of 7MW. Furthermore, the loss is categorized into steady-state component $P_{ss}$ and transient component $P_{ts}$ for separate statistical analysis, as shown in Fig. 13.

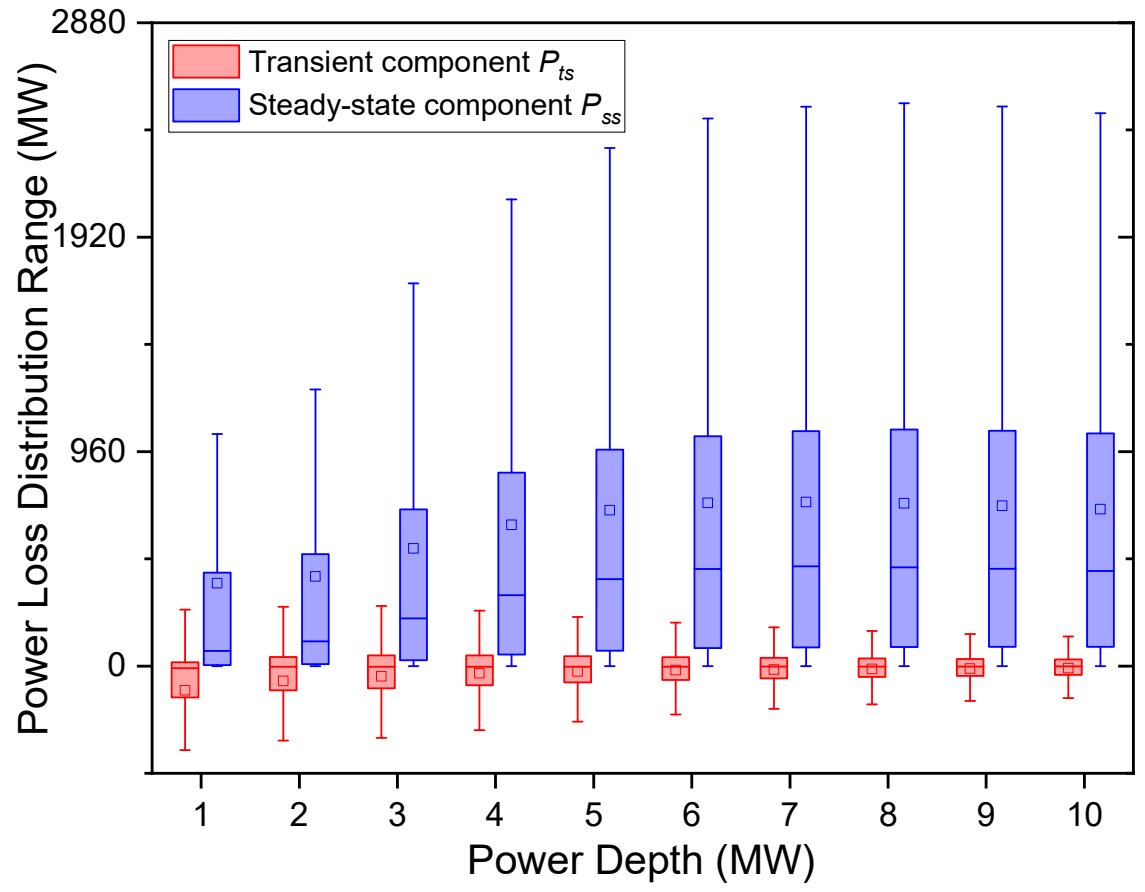

Fig. 13  Boxplot of $P_{ss}$ and $P_{ts}$ distribution

As the power depth increases, the distribution range of the steady-state component expands and eventually stabilizes, while the distribution range of the transient component narrows. Therefore, the proportion of the transient component in the total loss gradually decreases.

By combining the distribution patterns of $P_{clu}$ magnitude and $dP_{clu}/dt$, it can be observed that the trend of $P_{clu}$ magnitude distribution with respect to power depth is similar to the variation trend of the steady-state component $P_{ss}$, while the trend of $dP_{clu}/dt$ distribution with respect to power depth is similar to the variation trend of the transient component $P_{ts}$. This indicates that the overall magnitude of the input power $P_{clu}$ determines the size of the steady-state loss power component $P_{ss}$, while the short-term fluctuations of the input power are strongly correlated with the transient loss power component $P_{ts}$ of the battery.

## 3.4 The variation of battery loss energy $E_{loss}$.

By integrating the steady-state component $P_{ss}$ and the transient component $P_{ts}$ of the loss power over the entire time domain (one year), the steady-state loss energy $E_{ss}$ and transient loss energy $E_{ts}$ of the battery are obtained. The proportion of each is then statistically analyzed for power depths from 1MW to 10MW. The comparison results are shown in Fig. 14.

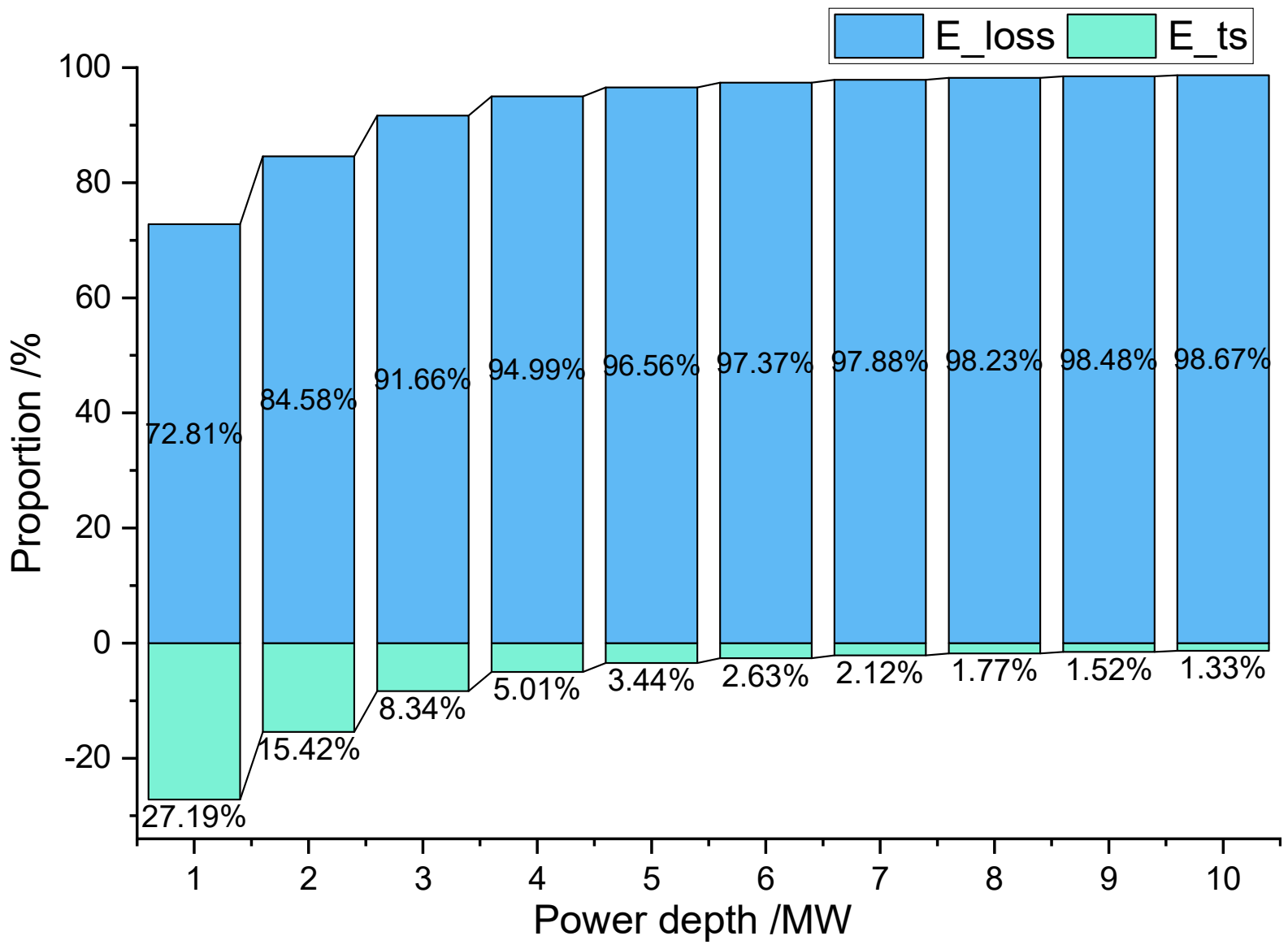

Fig. 14 Comparison between steady-state and transient components at different power depths.

In Fig. 14, the total length of the bar represents the steady-state loss energy $E_{ss}$, which is used as the reference value at each depth and displayed with a length of 100%. The positive half-axis indicates the actual battery loss energy $E_{loss}$, while the negative half-axis represents the integrated energy of the transient loss component $E_{ts}$, which is negative across all depths, reflecting the reduction in total battery loss due to polarization effects. In other words, if the battery model does not account for the impact of polarization reactions on loss, the total battery loss would be equal to the steady-state loss energy. Consequently, the battery loss energy calculated by the model would be higher than the actual value.

As the power depth increases, the proportion of $E_{ts}$ gradually decreases. At a depth of 1MW, $E_{ts}$ has the highest proportion, reducing the steady-state loss energy by 27.2%, meaning the total battery loss energy accounts for only 72.8% of the steady-state loss $E_{ss}$. However, when the power depth increases to 10MW, $E_{ts}$ reduces the steady-state loss energy by only 1.3%. With increasing power depth, the proportion of the base load rises while the share of fluctuating components decreases.

Based on the theoretical analysis of the transient component, the reduction in fluctuations leads to a diminished impact of the transient component.

The above data indicate that ignoring the impact of transient loss energy $E_{ts}$ on the overall battery loss is inappropriate when calculating battery loss.

### 3.5 Proportion of energy loss in different ESS components with varying power depth

Based on the previously described simulation model, this section presents the loss distribution under the current case conditions. The focus of this study is to analyze the energy consumption patterns of major power components within ESS. The energy consumption of non-power auxiliary components in the ESS, such as control, monitoring, and thermal management systems, is not included in the simulation.

According to the classification of model components, the primary sources of energy loss in the ESS include the transformer, AC/DC PCS, DC/DC PCS, battery ohmic resistance, and polarization resistance—five key components. The parameters of these five components have been introduced in previous sections. By categorizing based on components, the annual energy loss data is collected, and a comparison across power depths from 1MW to 10MW is conducted, ultimately forming the energy consumption proportion diagram for each component.

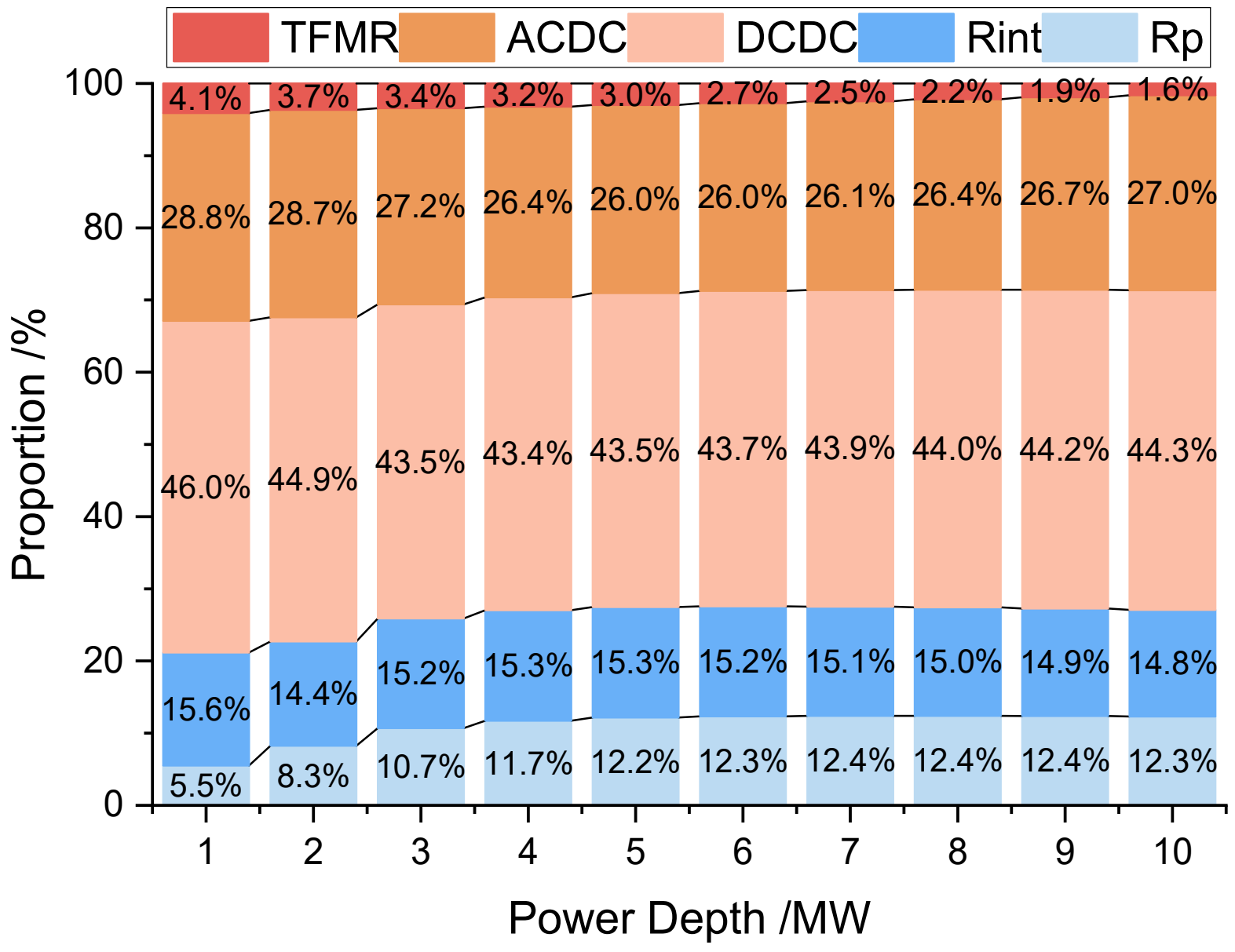

Fig. 15 Proportion of loss composition at different power depths

The statistical bars illustrates the loss proportions at different power depths. As the power depth increases, the proportion of battery loss slightly increases. After the power depth reaches approximately 7MW, the proportions of each component gradually stabilize.

## 4 Loss optimization method

In previous studies, the time-varying and time-invariant characteristics of BESS losses have been discussed, and corresponding energy storage optimization strategies have been proposed. This study will explore the input-efficiency equivalent relationship of BESS to perform instantaneous optimization of system losses at each time step. The BESS model in this study is built based on the simulation system presented in Section 1.1.

## 4.1 BESS Input-Efficiency statistics and equivalent relationship

BESS is connected to the low-voltage side of the transformer via an AC/DC PCS and is integrated into the grid after voltage step-up. Therefore, the interface variable between BESS and the grid is the charging/discharging power, which can be considered the sole input variable of BESS. Furthermore, for the BESS loss model, the primary focus is on BESS efficiency. Thus, to directly observe the loss characteristics of BESS, it is essential to explore the relationship between its input power and efficiency.

### 4.1.1 Statistical patterns of system power-efficiency

This study simulates a one-year BESS peak shaving scenario for a specific power grid based on its historical annual load, leveraging the peak shaving strategy in Section 2 and the BESS simulation system.

The BESS simulation system described in Section 1.1 can simulate the basic transient characteristics of the battery at both minute-level and hour-level time scales. The converter parameters are configured according to the Hongsheng PWS1 50k PCS, while the transformer parameters are set based on the reference values provided by the Chinese National Standard GB/T 6451-2023. The simulation takes the peak shaving scenario of a sample grid over a year as the case data source, recording power-efficiency operating points during the process and plotting them as a scatter plot of efficiency operating points, as shown in Fig. 16.

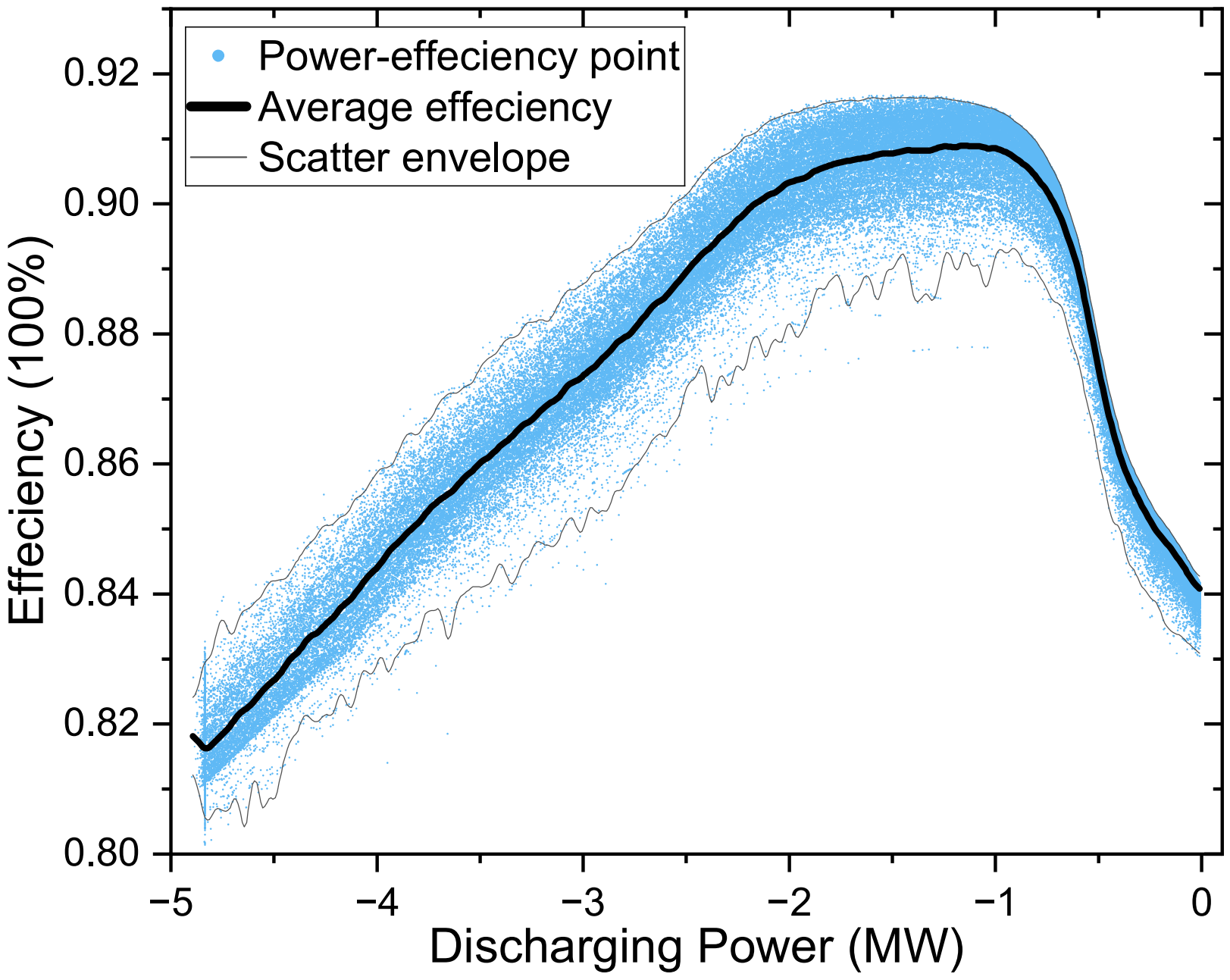

Fig. 16 Scatter plot of system discharging power-efficiency with median curve.

As shown in Fig. 16, the blue scatter points represent the power values on the horizontal axis and the total efficiency of the BESS on the vertical axis. The distribution range is enclosed by an envelope. The black line in the middle represents the median efficiency values for each discharge power.

The scatter exhibits a banded distribution, indicating that efficiency is not solely a function of input power. This is because the BESS model incorporates both minute-level and hourly transient characteristics of the battery. As a result, the efficiency of the battery is influenced by both the current magnitude and the temporal integration of the current, rather than being dependent on a single variable such as input power. Nevertheless, within certain accuracy constraints, the input-output relationship can be effectively approximated by the median line presented in Fig. 16.

Fig. 16 represents the discharge process of the BESS, while Fig. 17 depicts the charging process under the same plotting rules. The scatter distribution of power-efficiency during the charging process is comparatively more compact than that of the discharge process, as shown in Fig. 17.

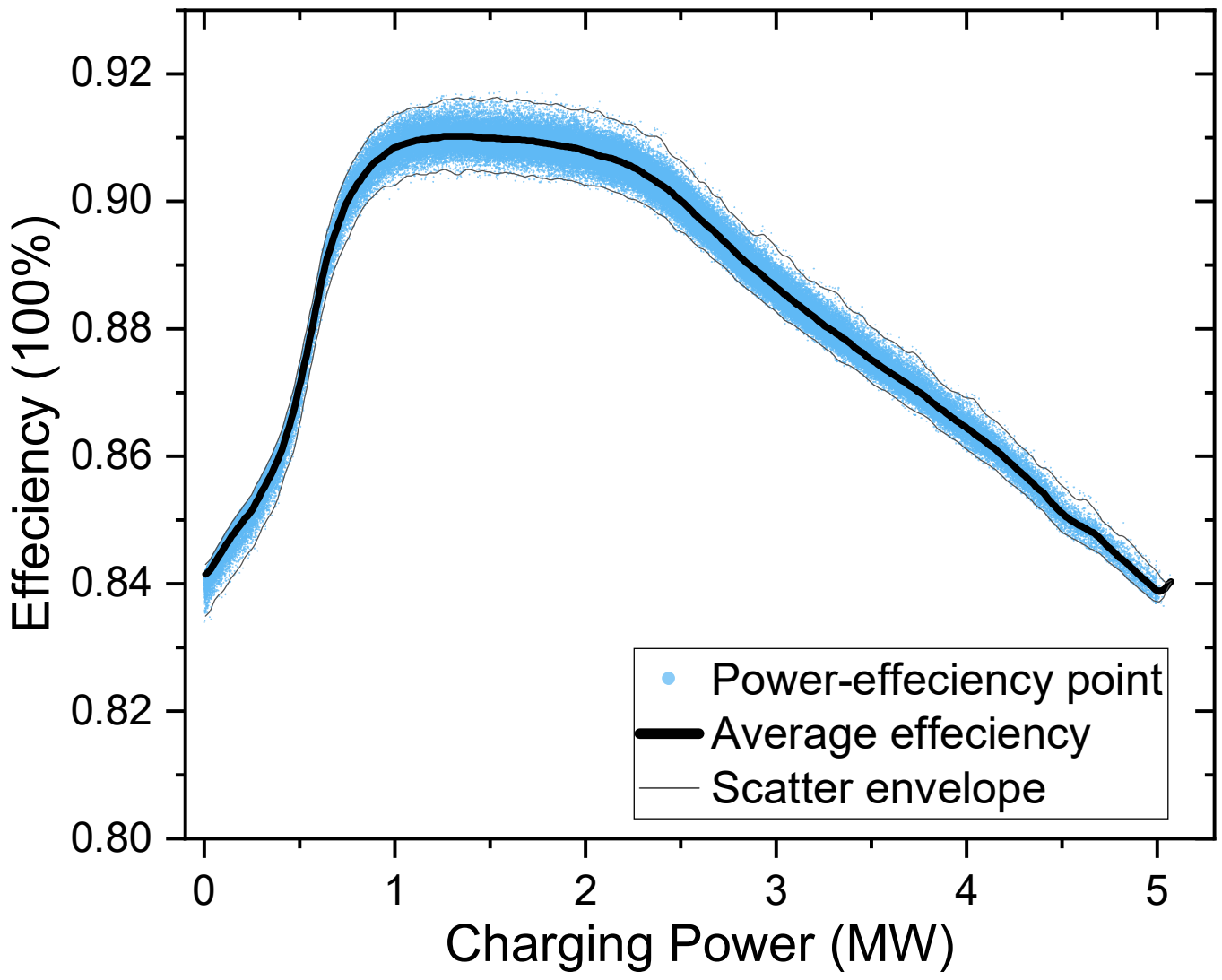

Fig. 17 Scatter plot of system charging power-efficiency with median curve.

## 4.1.2 Instantaneous loss optimization concept for bess

In summary, based on the aforementioned theory and previous research, the battery is inherently a time-varying system with respect to input power. However, the BESS input-efficiency data indicate that it can be approximated as a time-invariant system under certain conditions. Given the need for rapid or real-time optimization, this study neglects some time-varying characteristics, models the BESS as a time-invariant system with respect to input power.

If the Particle Swarm Optimization (PSO) algorithm is applied to optimize the power allocation of BESS battery clusters for maximum efficiency, the process can be essentially described as searching for the optimal operating point combination on the efficiency curves of each battery cluster at time *t*. This process is illustrated in the optimization schematic shown in Fig. 18.

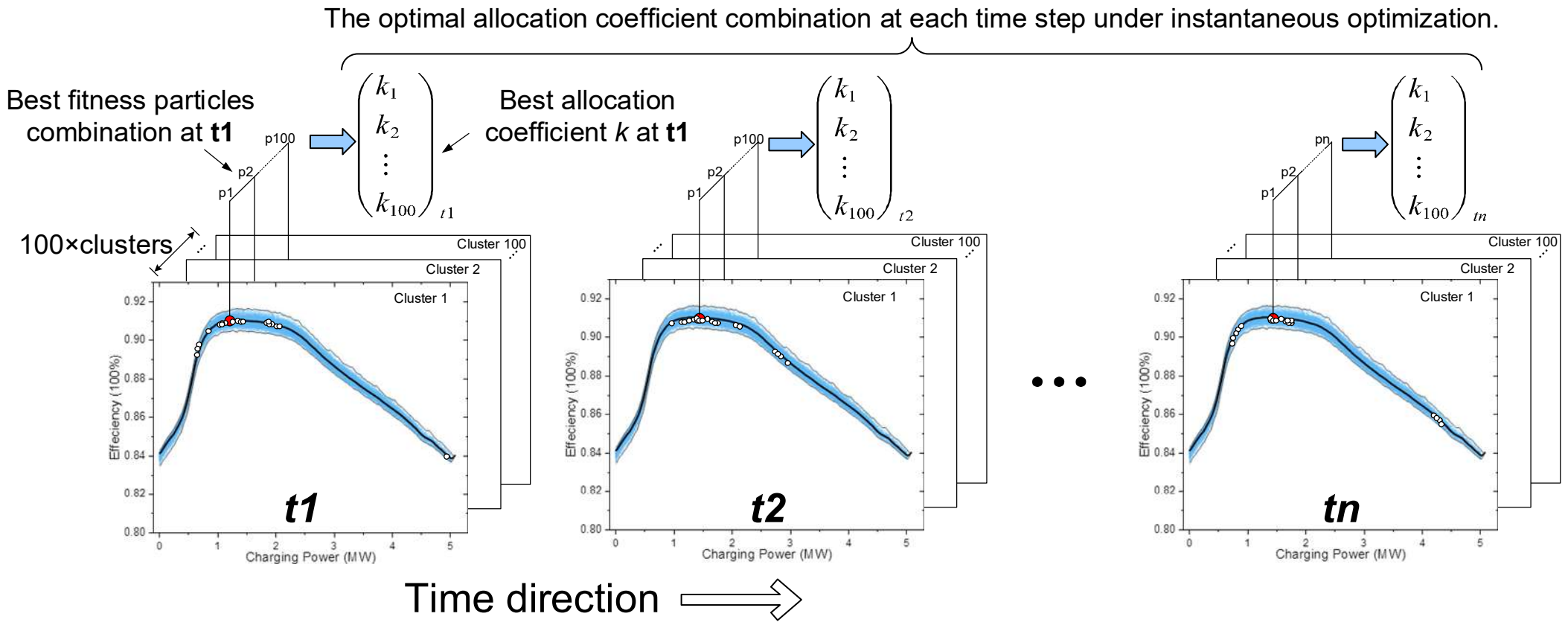

Fig. 18 Schematic diagram of the optimization process of PSO algorithm.

Based on this concept, this study proposes the following instantaneous loss optimization method for BESS.

## 4.2 BESS instantaneous efficiency optimization based on modified PSO algorithm

(1) Objective function

The overall loss optimization of BESS aims to maximize battery storage energy, and its objective function is as follows:

$$\max_{k_j} \ E(k_j, P_{sys}) = \sum_{j=1}^{m} k_j P_{sys}(t)\eta_{PCS.j}(t) - I_j^2(t)R_\Omega - \frac{I_j(t)}{C}\int_0^t I_j(s)e^{\frac{s-t}{R_pC}}ds \tag{14}$$

$K_j$ is the $j$-th battery cluster power allocation coefficient.

The PCS efficiency can be expressed as:

$$\eta_{PCS}(\lambda) = a_0 + a_1\beta + a_2\beta^2 + a_3\beta^3 + a_4\beta^4 \tag{15}$$

Definition of Load Rate $\beta$:

$$\beta \triangleq \frac{P}{P_N}, \ \beta \in [0, 1] \tag{16}$$

(2) Constraints

The constraint conditions include the constraint relationship between the total power and the battery cluster distribution coefficient:

$$P_{sys}(t) = \sum_{j=1}^{m} P_j(t) = \sum_{j=1}^{m} k_j P_{sys}(t) \tag{17}$$

Battery cluster power constraint relationship:

$$k_j P_{sys}(t)\eta_{PCS.j}(t) = I_j(t)\left[V_j(SoC) + I_j(t)R_\Omega + \frac{1}{C}\int_0^t I_j(s)e^{\frac{s-t}{R_pC}}ds\right] \tag{18}$$

Constraints on the value range of each battery cluster:

$$-P_N \leqslant P_j(t) \leqslant P_N \tag{19}$$

Energy constraints of each battery cluster：

$$P_j(t) = \begin{cases} k_j \times P_{sys}(t), \ k_j \in [0, 1] \ \ other\ condition \\ 0 \quad P_j(t) > 0 \ \& \ SoC(t) = SoC_{\max} \\ 0 \quad P_j(t) < 0 \ \& \ SoC(t) = SoC_{\min} \end{cases} \tag{20}$$

(3) Basic parameter settings for PSO

The initial particle values and fitness values are determined by the allocation coefficients in balanced mode and the battery energy increment. To balance performance and computation time, the particle count is set to 30, with a maximum iteration limit of 50.

The position update equation as follow:

$$k_{i+1} = k_i + \theta v_i + l_1 r_1(k_m - k_i) + l_2 r_2(k_{m_glb} - k_i) \tag{21}$$

In the improved PSO algorithm, the inertia weight is set to $\theta$=0.85, with the cognitive learning factor $l_1$=0.4 and the social learning factor $l_2$=0.5. Each particle maintains its historical best value $k_m$, while the global best value across all particles is denoted as $k_{m_glb}$. For the $i$-th iteration, the particle, representing an allocation coefficient, is denoted as $k_i$. To ensure solution stability and feasibility, the update velocity is constrained within $v_i \in [-1,1]$, and the particle values adhere to the allocation coefficient range $k_i \in [0,1]$.

The fitness function *f* in the PSO algorithm requires referencing the Simscape simulation model. To optimize computation time per iteration, two key methods are applied: FastRestart and Final/Initial states methods. FastRestart speeds up system resumption, cutting restart-related time and resources. The Final/Initial states methods define the system's start and end conditions. Together, they boost system simulation efficiency. Additionally, Parsim is used for parallel simulation to further reduce the computation time of the Simscape model. The allocation coefficient $k_{i+1}$ obtained from the PSO computation is first fed into the simulation model. Then, the simulation model returns the update of the fitness function *f*.

## 4.3 Case study

The optimized power allocation among battery clusters is applied to the peak shaving power demand $P_{dem}$ of a representative day. The resulting time-series data after optimization is shown in Fig. 19.

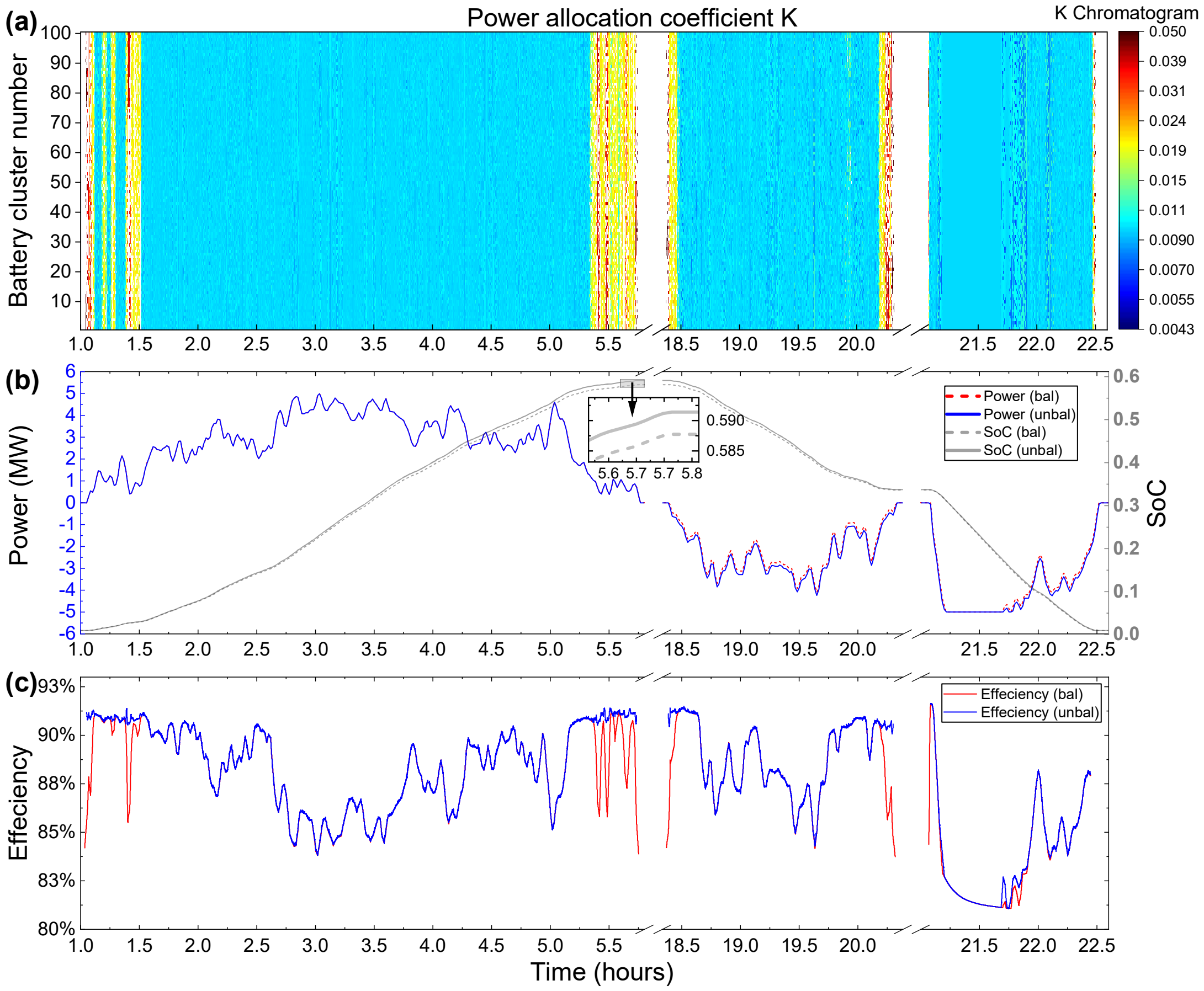

Fig. 19 Power allocation optimization result. (a) allocation coefficient. (b) power curve. (c) efficiency curve.

As shown in Fig. 19, the recorded data includes the 24-hour power allocation coefficients (heatmap), BESS charge/discharge power, SoC, and efficiency curves before and after optimization. The results indicate that, to achieve overall system efficiency optimization, battery clusters operate in an unbalanced state. While maintaining nearly the same total charge/discharge power, the optimized BESS achieves higher efficiency in specific time periods. Consequently, the maximum SoC value throughout the day is higher in the improved method compared to the original method.

As a control group, the energy loss data of each BESS component under the balanced power allocation mode, without loss optimization, was recorded over a day, as shown in Table 3.

Table 3 Loss and efficiency under original balance allocation method

| Items | Energy loss (kWh) | Proportion (%) | Average power efficiency (%) |
|---|---|---|---|
| TFMR | 156.31 | 3.32 | 99.54 |
| AC/DC | 1239.96 | 26.34 | 96.59 |
| DC/DC | 2104.66 | 44.71 | 94.24 |
| BAT | 1206.35 | 25.63 | 96.41 |
| Total | 4707.28 | 100 | 87.35 |

After the optimization of PSO, the power distribution coefficient of each battery cluster is instantaneously optimized. As shown in Table 4, the loss energy data of each component after optimization and the change of ratio are recorded.

Table 4 Loss and efficiency under optimized power allocation method

| Items | Energy loss (kWh) | | Proportion (%) | | Average power efficiency (%) | |
|---|---|---|---|---|---|---|
| | Value | Change | Value | Change | Value | Change |
| TFMR | 155.99 | -0.31 | 3.44 | +0.12 | 99.54 | 0.00 |
| AC/DC | 1167.25 | -72.70 | 25.75 | -0.59 | 96.59 | 0.19 |
| DC/DC | 1987.92 | -116.74 | 43.85 | -0.86 | 94.24 | 0.31 |
| BAT | 1221.90 | 15.55 | 26.96 | +1.33 | 96.41 | -0.06 |
| Total | 4533.07 | -174.21 | 100 | 0 | 87.35 | 0.40 |

From the data, the loss in one day after optimization is reduced by 174.21kWh, while the efficiency is improved by 0.4%. This shows the effectiveness of the power allocation strategy for BESS loss optimization. This verifies the effectiveness of the method.

# 6 Conclusion

By optimizing the peak shaving method to cycle-based control, the proposed strategy significantly enhances load energy capture (CR) and release rates (RR), achieving improvements from 1.0656 to 0.9978 and 2.1306 to 2.0528, respectively. This leads to a higher capacity utilization rate (CUR), increasing from 75.1% to 79.9%, and reduces equivalent cycle times by 7.83 cycles annually, thereby extending the lifespan of the Battery Energy Storage System (BESS).

Building on this, the study employs Particle Swarm Optimization (PSO) to optimize power allocation among battery clusters. This approach achieves a daily energy loss reduction of 174.21 kWh (3.7%) and increases system efficiency by 0.4%. The optimized power allocation enhances key performance indicators, including power output stability, State of Charge (SOC) consistency, and instantaneous efficiency, with SOC curves showing a more balanced and sustainable use of battery capacity.

Overall, the proposed strategy minimizes energy losses, enhances efficiency, and prolongs the operational lifespan of BESS, making it particularly effective for systems with high renewable energy penetration. These findings offer a practical framework for sustainable and cost-effective energy storage operations. Future work could refine this approach for more complex real-world scenarios, such as multi-cluster systems or dynamic grid conditions.